    \documentstyle[proceedings,numreferences,psfig]{maxent95} 

\begin{opening}
\title{DEPTH PROFILE RECONSTRUCTION FROM RUTHERFORD \protect\\ BACKSCATTERING DATA}

\author{U. v. Toussaint, K. Krieger, R. Fischer, V. Dose}
\institute{Max-Planck-Institut f\"ur Plasmaphysik, EURATOM Association\\
           POB 1533, D-85740 Garching, Germany 
	   \footnote{Email: udo.v.toussaint@ipp.mpg.de}}

\end{opening}

\runningtitle{DEPTH PROFILE RECONSTRUCTION FROM RBS-DATA}

\begin{document}

  \begin{abstract}
    An adaptive kernel method in the Bayesian framework together with a 
    new simulation program for
    Rutherford backscattering spectroscopy (RBS) have been applied to the analysis of 
    RBS data.
    Even in the case of strongly overlapping RBS peaks a depth
    profile reconstruction without noise fitting has been achieved. 
    The adaptive kernel method leads to the simplest depth profile
    consistent with the data. Erosion and
    redeposition rates of carbon divertor plates in the fusion experiment
    ASDEX Upgrade could be determined by RBS-analysis of thin film probes
    before and after exposition to plasma discharges.

  \end{abstract}

\section{Introduction}

Rutherford backscattering is one of the most important and most 
commonly applied techniques in surface 
analysis. Its main advantages are that it is fully quantitative 
and that precisions less  
than 1\% can be achieved \cite{Jeynes97}. The interpretation of the data, however, is in many 
cases not
straightforward. During the last decade several computer programs for the simulation and 
analysis of spectra obtained from RBS were developed, such as RUMP \cite{Doolittle85} 
or SIMNRA \cite{Mayer97}.
With these programs the determination of a depth profile is, however, a matter of 
trial and error.
The user has to prescribe depth profiles of all 
elements and has to compare the simulated
spectrum calculated from the input profiles with the data. 
The depth profiles are then adjusted until one obtains a reasonable agreement of
simulated and measured data.
Obviously this evaluation procedure has several shortcomings. 
It is a time-consuming cumbersome task,
the accuracy of the achieved depth profile is unknown and in many cases there is an ambiguity
between different depth profiles which fit the data equally well. 
The combination of the adaptive kernel method in the Bayesian framework \cite{Fischer97} 
with an RBS-simulation 
program allows to
overcome these disadvantages and extends the 
potential of Rutherford backscattering spectroscopy.

\section{Basic Concepts of Rutherford Backscattering}

In RBS-analysis, a sample is exposed to  a beam of ions with mass $m_{0}$ 
(e.g. He-particles) with a well defined energy $E_{0}$
in the order of MeV. Ions undergoing elastic Coulomb 
collisions with sample atoms
are recorded in a solid state detector
which views at a fixed deflection angle $\theta$. The Rutherford cross-section for 
this coulombic projectile-target interaction is 
quantitatively known. The energy $E'$ of the backscattered ions depends 
on the energy $E$ before
the collision, the mass of the ions $m_{0}$, the mass of their colliding 
partner $M_{i}$ and the 
deflection angle $\theta$ :
\begin{equation}
E'=E\left[\frac{\sqrt{1-\left(\frac{m_{0}}{M_{i}}\right)^{2}\sin^{2}\theta}+\frac{m_{0}}{M_{i}}\cos\theta}{1+\frac{m_{0}}{M_{i}}}\right]^{2}.
\end{equation}
From Eq. 1 we see that ions undergoing a collision with a heavy target atom 
loose less
energy than ions colliding with a target atom of lower atomic mass.
In addition, both primary ions and scattered ions loose energy on their way through the sample, 
depending on the
stopping power. This is the main reason which enables RBS to be depth sensitive.
The stopping power depends on the energy of the particles and the composition of the sample.\\
\begin{figure}[htb]
\centerline{\psfig{file=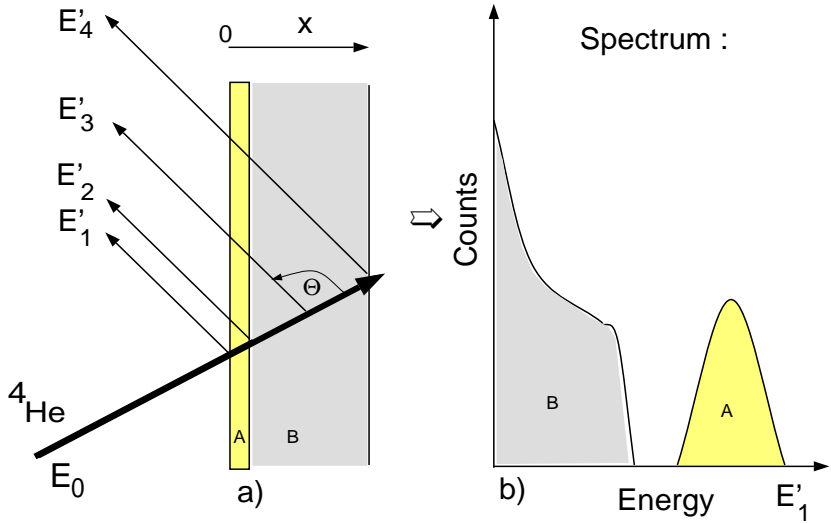,width=11cm,height=7cm}}
\caption{\textit{Schematic diagram of a RBS-experiment a) and the corresponding spectrum b).}}
\label{fig.RBS_Schematic}
\end{figure}
Fig. {\ref{fig.RBS_Schematic}a} depicts a typical RBS experiment. A thin overlayer
(A) of atoms with a high atomic mass $M_{A}$ is on top of the bulk substrate (B) with a
lower atomic mass $M_{B}$. In the energy spectrum of backscattered 
particles (Fig. {\ref{fig.RBS_Schematic}b}), the
film A leads to a spectral peak at higher energies, broadened by the apparatus 
transfer function
and the statistical fluctuations of the energy 
loss of the ions.
Scattering from B produces a broadened step at lower energies. 
The high energy side of this step 
originates from 
scattering from the topmost B-Layer. The increase of the spectrum with decreasing energy 
results mainly from the $\frac{1}{E^{2}}$
dependence of the Rutherford cross section.

\section{Simulation of RBS-Spectra}

For a spectrum synthesis the sample is divided into sub-layers with thickness $\Delta x$.
The spectrum is calculated from the superimposed contributions of scattering processes
from all elements in all sub-layers of the sample. 
For each sub-layer the concentrations on the layer-boundaries must be given. Inside
the sub-layer the concentration profile is assumed to interpolate linearly.
In each sub-layer the energy loss of the ions inside this layer and the 
cross-sections are determined.
\paragraph{Cross-Section Data:}
The actual cross-section deviates from the well known \\
Rutherford cross-section \cite{Tesmer95} at
both, high and low energies. The low-energy discrepancy is caused by partial screening of the
nuclear charges by the electronic shells \cite{Tesmer95}. This screening is taken into 
account by a
correction factor $C(E,\Theta)$ \cite{Anderson80}.
At high energies the cross sections deviate from the Rutherford cross-section due to the 
influence of the nuclear force \cite{Bozoin95}. This is unimportant in the present case.
\paragraph{Stopping Power Data:}
The two dominant processes of energy loss of a penetrating ion are the interactions of the
moving ion with bound or free electrons in the target, and the interactions of the moving 
ion with the screened or unscreened nuclei of the target atoms.
The electronic stopping power data are taken from Ziegler, Biersack and Littmark
\cite{Ziegler85}.
The nuclear stopping power for helium is calculated 
from \cite{Ziegler85}.
In compound materials, Bragg's rule is used,
\begin{equation}
{\left(\frac{dE}{dx}\right)}_{total}=\sum_{i}c_{i}{\left(\frac{dE}{dx}\right)}_{i},
\end{equation}
to calculate the effective stopping power ${\left(\frac{dE}{dx}\right)}_{total}$ from the concentrations $c_{i}$
and the stopping power ${\left(\frac{dE}{dx}\right)}_{i}$ of each individual component $i$.
The key assumption of Bragg's rule that the interaction between the ion and a target atom is
independent of the environment holds in most cases. In some compounds such as
oxides the deviations from Bragg's rule predictions may, however, be of the order of 
10\% to 20\% \cite{Ziegler88}.
\paragraph{Energy Loss Straggling:}
The energy loss of charged particles penetrating material is accompanied by a spread of the 
beam energy which is due to statistical fluctuations of the energy transfer 
in the loss channels. 
As the number of interactions is high, the energy broadening is well described by a  
Gaussian. The program uses Bohr's theory of energy-loss straggling \cite{Bohr48}, together with
corrections by Chu \cite{Chu76}, which include the electron binding in the target atoms.
The energy dependence of the stopping power results further in a non-stochastic 
broadening (or squeezing)
of the energy distribution of the ion beam. The energy width $\Delta E_{f}$ after passing the
sub-layer is given by \cite{Szilagy95}:
\begin{equation}
\Delta E_{f}=\frac{S(E_{f})}{S(E_{i})}\Delta E_{i}
\end{equation}
with $E_{i}$, $E_{f}$ as the mean energies and $S(E_{i})$, $S(E_{f})$ as the stopping powers at 
the entrance and exit of the sub-layer, respectively.

\section{Experiment}

\begin{figure}[htb]
\centerline{\psfig{file=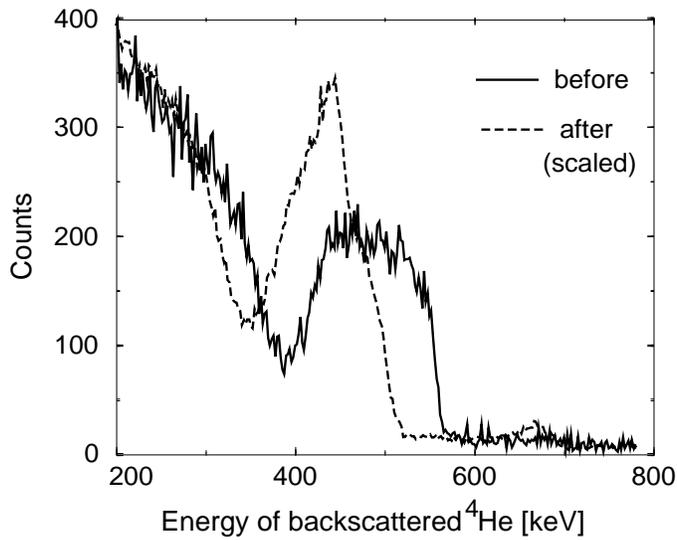,width=9cm}}
\caption{\textit{RBS-spectra before and after plasma exposure. The shift of the high
energy edge is clearly visible.}}
\label{fig.RBS_Daten}
\end{figure}

The interpretation of RBS data is required for the analysis of erosion 
measurements of plasma facing materials in fusion experiments.
The solid inner walls surrounding the plasma are
subjected to an intense bombardment by plasma particles because the confinement of 
the plasma by the confining
magnetic field is not perfect. The surfaces of the inner walls are mainly modified 
by ion implantation, erosion and
by deposition of material from other wall areas. \\
One major problem in fusion research is to
find a wall material where wall erosion rate and wall modifications are small and 
tolerable \cite{Behrisch88}.
The importance of this problem for planned fusion power plants is emphasized by an erosion 
analysis
for ITER \cite{Brooks98}. The modeled gross erosion yield of a carbon-divertor could reach 
a maximum of $5$m/burning-year, which is reduced
by redeposition down to about $0.5$m/burning-year. The modeling, however, faces 
exceptional difficulties
due to complex hydrocarbon transport phenomena and the lack of 
input data (e.g. for low energy sputtering).
Therefore experimental determination of erosion and redeposition yields 
is necessary to
validate the modeling and to improve the quantitative knowledge of the fundamental 
erosion processes.\\
To determine carbon erosion rates in the divertor of ASDEX Upgrade, graphite probes
which were covered with a $150$nm layer of $^{13}C$ were exposed to single plasma discharges.
$^{13}C$ was used because chemical erosion is unaffected by isotope
substitution and to allow the measurement of redeposited $^{12}C$ eroded at other
plasma facing components.
Furthermore the stopping power in $^{13}C$ and $^{12}C$
is the same and so the
limited accuracy of the stopping power in the simulation cancels.
\begin{figure}[htb]
\centerline{\psfig{file=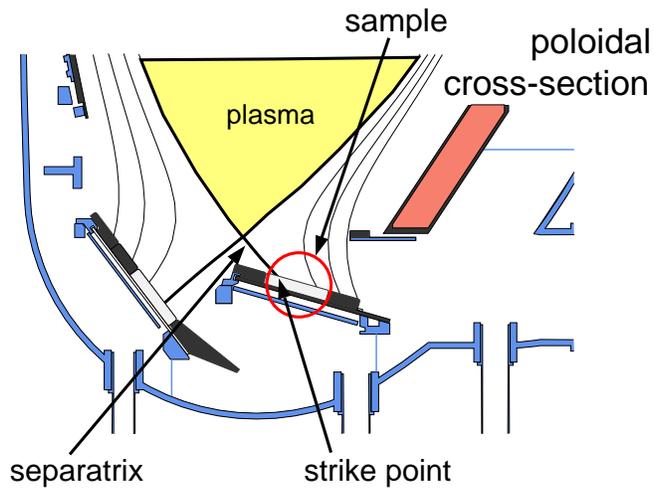,width=9cm}}
\caption{\textit{Poloidal cross-section of ASDEX-Upgrade. The circle 
indicates the position of the sample on the outer divertor 
in ASDEX-Upgrade. 
The separatrix is the outermost closed magnetic flux line. The point 
the separatrix touches the divertor is called the strike point}}
\label{fig.Divertor}
\end{figure}
The sample was introduced in the outer divertor of ASDEX Upgrade 
(circle in Fig. {\ref{fig.Divertor}})
covering in particular the  strike point, which is the point where the outermost 
last closed magnetic 
flux line touches the plate surface with a corresponding maximum of the power load.
\\
The samples were analyzed before and after plasma exposure with a total 
exposure time of 4 seconds using RBS with 2.0 MeV $^{4}$He ions.
The backscattered particles were detected at a scattering angle
of $\Theta=165^{\circ{}}$. The width of the apparatus transfer function 
is about 19keV FWHM \cite{Dose97}.
Fig. {\ref{fig.RBS_Daten}} shows typical spectra before and after plasma exposure.
Before plasma exposure the signal from the $^{13}C$-layer at higher energy is separated by a 
gap from the part of the spectrum corresponding to the underlying $^{12}C$-bulk material.
After plasma exposure the high energy edge of the signal from $^{13}C$ has shifted 
towards lower energies. This indicates that there is no 
longer $^{13}C$ at
the surface of the sample. The peak at $430$ keV is due to 
the $^{12}C$ at the sample surface and from the $^{13}C$ fraction below the surface.
The difference of the RBS-spectra before and after exposure contains the information
about the erosion and redeposition yields.

\section{Results}

\begin{figure}[htb]
\centerline{\psfig{file=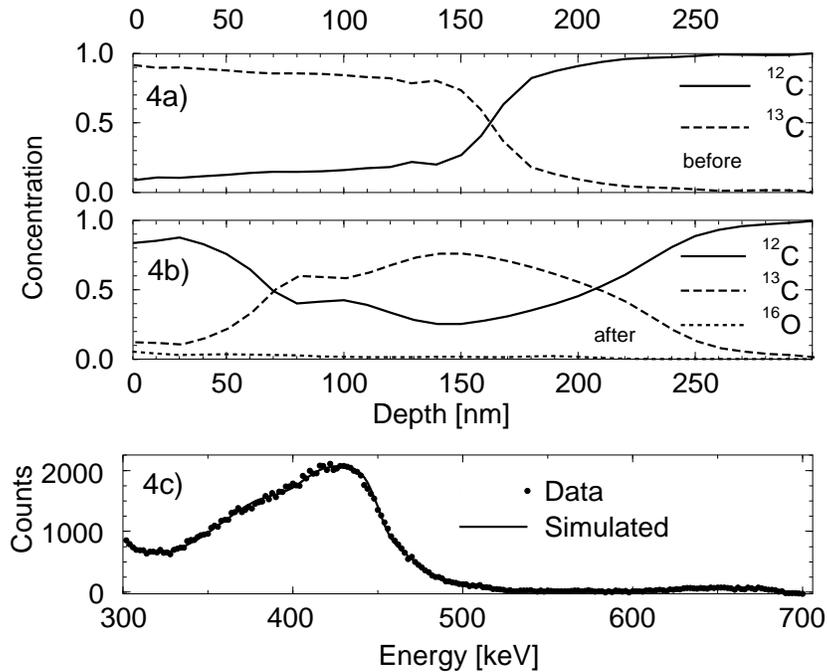,width=11cm}}
\caption{\textit{panels a) and b): $^{12}C$ and $^{13}C$-distribution before 
and after plasma exposure. 
Panel c): RBS-data 
(black dots) and the calculated RBS-spectrum 
(grey line) from the depth profile in the panel b).}}
\label{fig.Depth_Profiles}
\end{figure}

\begin{figure}[htb]
\centerline{\psfig{file=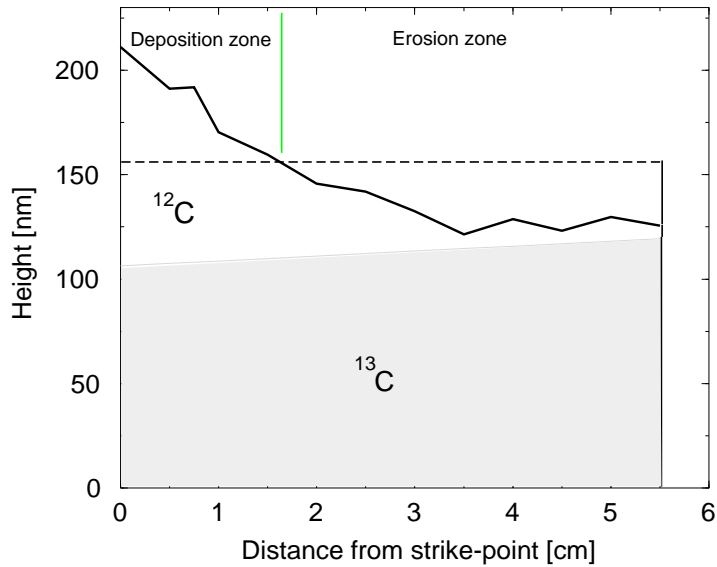,width=9.5cm}}
\caption{\textit{Schematic picture of the $^{12}C$ and $^{13}C$ distribution before and
after plasma exposure. The grey dashed line gives the height of the $^{13}C$-tracer
before plasma exposure. The grey shaded area marks the height of $^{13}C$ after plasma
exposure and the difference between the upper black line and the grey shaded area gives the
height of deposited $^{12}C$.}}
\label{fig.Result}
\end{figure}

To determine the concentration depth profiles from the measured RBS data a simple
$\chi^{2}$-fit is insufficient and results in useless rapidly oscillating depth profiles.
This is due to the ill-conditioned nature of the inversion problem which results from the
energy-straggling broadening, the finite apparatus-induced energy resolution and the 
counting statistics.
Furthermore the optimal grid, given by the thickness of the sub-layers 
the sample is divided in,
is unknown. \\
For this kind of problems the adaptive kernel method is well suited. 
The concept of adaptive kernels provides local smoothness which makes the
result robust against noise corruption. The locality of the information content of
the data is taken into consideration by the local varying kernel widths.
Constraints
like positivity or other prior knowledge (like bulk concentrations) are easy to include.
The used adaptive kernel method is presented in detail in this proceeding \cite{Rainer99}.
\\Fig. {\ref{fig.Depth_Profiles}a} shows the reconstructed $^{12}C$ 
and $^{13}C-$depth
profiles of a sample before plasma exposure. The concentrations in each layer sum up to one.
The surface concentration of $^{13}C$
(on the left side) is above 90\% and decreases only slightly to a depth of about $150$nm.
The remaining 10\% fraction of $^{12}C$ is caused by impurities in the coating process. 
The broad transition between the $^{13}C-$layer and the $^{12}C-$bulk can be explained 
by the interface roughness of the virgin sample. 
After 4 seconds of plasma exposure the depth profiles have changed 
dramatically, as shown in Fig. {\ref{fig.Depth_Profiles}b}. 
There is a $^{12}C-$layer with a thickness of about
$70$nm on top of the $^{13}C$. The maximum concentration of $^{13}C$ has decreased, however, 
the thickness of the $^{13}C$-layer is with about $170$nm nearly unchanged. 
Furthermore, there is a continuous
level of $^{12}C$ in the whole sample with a minimum concentration of 20\%.
Since diffusion due to thermal effects could be excluded, the impacting $^{12}C$-atoms must 
have mixed the material.
Fig. {\ref{fig.Depth_Profiles}c} shows the RBS-data
as black dots and the calculated RBS-spectrum (solid line) based on the depth profile shown 
in Fig. {\ref{fig.Depth_Profiles}b}. The
agreement is within the counting statistics.\\
With samples in different distances to the strike point we achieved a
laterally resolved determination of erosion and deposition 
as shown in Fig. {\ref{fig.Result}}. The height of the $^{13}C$-tracer
was $153$nm before exposure (dashed line in Fig. {\ref{fig.Result}}).
The grey shaded area marks the thickness of the $^{13}C$-layer after plasma exposure. The
highest erosion of $40$nm was observed at the strike point. 
With increasing distance the erosion 
reduces slightly to $\simeq{30}$nm in $5$cm distance. The solid line represents the
joint height of the $^{13}C$ and deposited $^{12}C$ under the assumption that no $^{12}C$
from the bulk was eroded. The difference between the solid line and the grey shaded
area of $^{13}C$ is the height of deposited $^{12}C$. The amount of $^{12}C$ which covers the
$^{13}C$ is largest at the strike point with over $100$nm and reduces down to $10$nm
in a distance of $5.5$cm.
Near the strike point the redeposition of carbon is larger than the erosion, 
which makes this location a net deposition zone. By contrast, in a distance larger 
than $1.5$cm from the strike 
point there is a net erosion area. Fig. {\ref{fig.Result}} is only a schematic representation
which shows the total amount of $^{12}C$ and $^{13}C$ in a simplified distribution.
It can be seen from the depth profiles in Fig. {\ref{fig.Depth_Profiles}} 
that after 
plasma exposure there are no longer clearly separated layers of the two different isotopes
and pronounced mixing has occurred. The large spatial variation of erosion and deposition
rates shows, that the lifetime of plasma facing components can only be evaluated
for specific local conditions.

\section{Conclusions}

With the used combination of the RBS-simulation program and the adaptive kernel method
the capabilities of RBS-data evaluation have been considerably extended.
This allows to study erosion, deposition and mixing of carbon as inner wall material
in fusion experiments by using different isotopes which have no influence on the 
chemical erosion.
The experiment shows a spatially varying net erosion/deposition rate with large mixing.
Further investigations are necessary to answer the question of the long-time behavior
of the erosion of the inner wall materials facing different plasma conditions.

\bibliography{article}
\bibliographystyle{maxent95}

\end{document}